\documentclass[12pt]{article}
\usepackage[cp1250]{inputenc}
\usepackage{amsmath}
\usepackage{amssymb}

\newtheorem{Theorem}{Theorem}[section]

\newtheorem{Lemma}[Theorem]{Lemma}
\newtheorem{Definition}[Theorem]{Definition}
\newenvironment{Proof}[1]{{\bf Proof #1.} }{$\Box$\\}

\newcommand{\vecDD}[2]{\left[\begin{array}{c} #1 \\ #2 \end{array}\right]}

\newcommand{\comment}[1]{}

\numberwithin{equation}{section}

\begin{document}

\centerline{\Large Directional approach to spatial structure of solutions\\}
\smallskip
\centerline{\Large to the Navier-Stokes equations in the plane}
\bigskip
\smallskip
\centerline{Pawe\l{} Konieczny${}^a$ and Piotr Bogus\l aw Mucha${}^b$}

\begin{center}
{${}^a$Institute for Mathematics and Its Applications}\\
{University of Minnesota}\\
{114 Lind Hall, 207 Church St. SE, Minneapolis, MN, 55455, U.S.A.}\\
{E-mail: \textit{konieczny@ima.umn.edu}}\\\smallskip
{${}^b$Institute of Applied Mathematics and Mechanics}\\
{Warsaw University}\\
{ul. Banacha 2, 02-097 Warszawa, Poland}\\
{E-mail: \textit{mucha@hydra.mimuw.edu.pl}}
\end{center}

{\bf Abstract.} 
We investigate a steady flow of incompressible fluid in the plane. The motion is governed by the Navier-Stokes equations
with prescribed velocity $u_\infty$ at infinity. The main result shows the existence of unique solutions for arbitrary
force, provided sufficient largeness of $u_\infty$. Furthermore a spacial structure of the solution is obtained in comparison
with the Oseen flow. A key element of our new approach is based on a setting which treats the directino of the flow as \emph{time}
direction. The analysis is done in framework of the Fourier transform taken in one (perpendicular) direction and a special choice
of function spaces which take into account the inhomogeneous character of the symbol of the Oseen system.
From that point of view our technique can be used as an effective tool in examining spatial asymptotics of solutions to other systems
modeled by elliptic equations.
\medskip


\section{Introduction}
% label_prefix:intr

We investigate the problem of existence and asymptotic behaviour of solutions to the stationary Navier-Stokes equations
in the full space:
\begin{eqnarray}
	u\cdot\nabla u - \Delta u + \nabla p & = & F \qquad \textrm{in~} \mathbb{R}^2,\label{ch4:intr0}\\
	\mathrm{div~}u & = & 0 \qquad \textrm{in~} \mathbb{R}^2,\label{ch4:intr10}
\end{eqnarray}
where $u$ is a sought velocity, $p$ a corresponding pressure and $F$ is a given external force. This system is coupled with a condition at infinity:
\begin{equation}\label{ch4:intr20}
	u \to u_\infty \qquad \textrm{as~}|x|\to\infty.
\end{equation}
This problem is connected with a question of a flow around an obstacle (that is when considered domain $\Omega$ is
$\Omega = \mathbb{R}^2 \setminus K$, where $K$ is a  compact domain) -- a problem which has a long history
but is far from being completely solved.
So far one can present the following main contributions to the problem. The first one is due to Leray (\cite{Leray}) where he proves
existence of a solution $u_L$ (Leray solution) with finite Dirichlet integral
\begin{equation}\label{uLDirichlet}
	\int_{\Omega} |\nabla u_L|^2 dx < \infty.
\end{equation}
The technique used there is to create a sequence of solutions in domains $\Omega_R = \Omega \cap \{ x : |x| < R \}$,
for which $\int_{\Omega_R} |\nabla u_R|^2 dx < C$, where $C$ is independent of $R$. After taking a suitable subsequence as $R \to \infty$
one obtains a solution $u_L$ to (\ref{ch4:intr0})-(\ref{ch4:intr10}). Unfortunately Leray was not able to prove that his solution
satisfies (\ref{ch4:intr20}) or even that it is not trivial. The reason for that is that the power $2$ coincides with a dimension $2$ in (\ref{uLDirichlet})
giving no information about the behaviour of $u_L$ at infinity. Thus a different approach was needed.

Regarding existence to the  N-S system with (\ref{ch4:intr20}) result of Finn and Smith \cite{FiSm} should be noted. Under smallness
assumptions on $u_\infty$ Authors solve the problem using perturbation approach and the contraction mapping principle. A similar approach
was also followed by Galdi \cite{Galdi1993}, where the Author used $L_p$ estimates for the Oseen system. Those estimates were also used
in \cite{GaSo1995} by Galdi and Sohr to show that under a suitable integrability condition the solution of the Navier-Stokes system is a PR solution.
It is important to mention that these
results strongly depend on the assumption that $u_\infty \neq 0$, that is in the case, where linearization of the Navier-Stokes equations
result in the Oseen and not the Stokes system. Solutions to the former have better properties at infinity due to the presence of the term $u_{,1}$.

The main problem is to show convergence to the prescribed velocity field at infinity. It was addressed by Gilbarg and Weinberger 
in two classical papers \cite{GiWa1}, \cite{GiWa2}. They investigate behaviour of Leray solutions $u_L$ and later more general of
all solutions with finite Dirichlet integral. Their approach uses structure of the Navier-Stokes equations, more precisely
equation for the vorticity $\omega = u_{,2} - u_{,1}$:
\begin{equation}
-\Delta \omega + u\cdot\nabla \omega = 0,
\end{equation}
for which one has a maximum principle for $\omega$ (a similar equation is also investigated for the total-head pressure $\Phi = p + \frac{1}{2}|u|^2$).
In particular Authors show that the Leray solution $u_L$ is bounded and there exists
a constant vector field $\widetilde{u_\infty}$ such that
\begin{equation}\label{uLConvergence}
 \int_0^{2\pi} |u_L(r, \theta) - \widetilde{u_\infty}| d\theta \to 0 \quad \textrm{as~} r\to\infty.
\end{equation}
In general case (that is not necessarily via Leray's approach) the same property holds if one assumes first that this solution is bounded.

Even though this result was a considerable step forward many questions remained open, for example if $\widetilde{u_\infty} = u_\infty$ or
if the convergence (\ref{uLConvergence}) can be improved.

The next important and classical result was due to Amick \cite{Amick}. In particular the author generalized the results by Gilbarg and Weinberger
by showing that any solution with bounded Dirichlet integral is in $L^\infty(\Omega)$. Also the convergence (\ref{uLConvergence})
is improved to the pointwise one but only for symmetric flows. In the same paper a solution due to Leray is investigated,
in particular it is shown that in case of symmetric flow this approach results in a non-trivial solution.

These mentioned results do not address the asymptotic behaviour of solutions, which itself is also an interesting problem. One of the most
classical papers regarding asymptotic profile of solutions is an article by Finn \cite{Finn}, where the Author introduced a class of solutions
(to the three-dimensional Navier-Stokes equations) called Physically Reasonable (\textit{PR}). The two dimensional case was covered
by Smith \cite{Smith}. The definition of \textit{PR} solutions in 2D is as follows: 
\begin{Definition}
 A solution $u(x)$ of (\ref{ch4:intr0})-(\ref{ch4:intr10}) is said to be \textit{PR} if there is a positive constant $C$, a positive number $\epsilon$,
and a vector field $u_0 \neq 0$ such that $|u(x) - u_0| < C|u_0 r|^{-\frac{1}{4}-\epsilon}$ as $|u_0 r|\to\infty$, where $r = |x|$. 
\end{Definition}
This class is called Physically Reasonable because these solutions have behaviour expected from the physical point of view, for example
existence of a wake region behind the obstacle and indeed a faster decay than assumed. In particular Smith showed that all these solutions
satisfy the following estimate $|u(x) - u_0| = O(|x|^{-1/2})$, which is sharp in the sense that only very specific flows satisfy $|u(x) - u_0| = o(|x|^{-1/2})$.

\smallskip
In this paper we consider the case of a full plane flow with a goal to develop new methods to deal
with a problem in exterior domain. 
Our techniques require non-standard function spaces due to the fact that our analysis is carried through in a Fourier space only in one direction. Of course, one may interpret these results in Sobolev spaces, but we do not address this question here. A hint to dicover our space setting one can find in considerations for the evolutionary Navier-Stokes in \cite{SiAr}, however this system is homogeneous which is a different situation than in the present paper.
Similar techniques were used by Wittwer in \cite{Wittwer1}, however
for a symmetric flow. In \cite{HaldiWittwer} authors ommited this condition, however
these techniques are still very technical. Symmetry condition was also needed in the recent work of Yamazaki \cite{Yamazaki} for the
same problem as ours.

Our result does not rely on any symmetry and is also much simpler then in \cite{Wittwer1} -- we do not consider
system for the rotation of the fluid, but we operate on the velocity vector field itself. Moreover,
we give analysis of the flow not only in the halfplane, but in the full plane omitting artificial boundary conditions. In such case
one can obtain not only asymptotic behaviour behind the obstacle, but also in front of it. 

The existence of the solution is also shown under assumption on the force $F$ in comparison to $u_\infty$ in this sense
that we show existence for arbitrary large force $F$ provided $u_\infty$ is large enough.
This is an improvement of result from \cite{Yamazaki} since we do not impose any symmetry for external force $F$.
Moreover our method allows us to obtain exact asymptotic profile of the solution -- the result which is typically shown by referring to
work of Smith \cite{Smith} (see \cite{Amick2}). In particular we obtain directly that our solution behaves like $O(|x_1|^{-1/2})$.

Although the Oseen system has better properties than the Stokes one, one difference must be emphasized -- its inhomogeneity. The term
$u_\infty\partial_{x_1} v$ in (\ref{Oseen10}), which sometimes improves information about the sough solution, causes that the Oseen system is not invariant under any rescalling. This follows that there is no natural regularity structure of the system. Hints to find an aprioprate function space can be found in our previous papers \cite{Ko,Mu}, where the phenomenon of the wake region as a consequence of a parabolic degeneration has been partly explained.

Throughout the paper we use the following Banach spaces:
\begin{Definition}
	Let $\beta \in \mathbb{R}$. 
	Space $\mathcal{X}_\beta$ consists of these functions, for which the following
	norm is finite:
	\begin{equation*}
		\|a\|_{\mathcal{X}_\beta} = \sup_{(t, \xi)\in \mathbb{R}^2}	(1+|t\xi^2|)^\beta |a(t, \xi)|	.
	\end{equation*}
	Function space $\mathcal{Y}$ consists of these functions, for which the following norm is finite:
	\begin{equation*}
		\|b\|_{\mathcal{Y}} = \sup_{(t, \xi)\in \mathbb{R}^2 } |t|^{1/2}|b(t, \xi)|.
	\end{equation*}
\end{Definition}

The technique used in this work is to consider parallel to $u_\infty$ coordinate as time and then use the Fourier transform
to obtain system of ordinary differential equations, which can be solved and analyzed to obtain
information about its asymptotic behaviour. 

Taking (\ref{ch4:intr20}) into account we may introduce a new vector field $u = u_\infty + v$, for which we have
the following system:
\begin{eqnarray}
	u_\infty v_{,1} + v\cdot\nabla v - \Delta v - \nabla p & = & F \qquad \textrm{in~} \mathbb{R}^2,\label{ch4:intr30}\\
	\mathrm{div~}v & = & 0 \qquad \textrm{in~} \mathbb{R}^2,\\
	v & \to & 0 \qquad \textrm{as~}|x|\to\infty.\label{ch4:intr50}
\end{eqnarray}

Our main result states the following:
\begin{Theorem}\label{ch4:mainTheorem}
	Considering system (\ref{ch4:intr30})-(\ref{ch4:intr50}). Given fixed constant $\beta \in (1/4, 1/2)$ and given $F$ and $u_\infty$ such that 
	$$\|\hat{F}/|\xi|\|_{\mathcal{Y}} \leq Cu_\infty^{2\beta-1/2},$$
	where the constant $C$ is independent of the data. Then there exists a unique solution $(v, p)$ to (\ref{ch4:intr30})-(\ref{ch4:intr50}) such
	that $\hat{v}\in \mathcal{X}_\beta$, $\hat p\in \mathcal{Y}$ and the following estimate is valid:
	\begin{equation*}
		\|\hat v\|_{\mathcal{X}_\beta} + \|\hat p\|_{\mathcal{Y}} \leq Cu_\infty^{2\beta - 1/2}.
	\end{equation*}
\end{Theorem}

The proof of this theorem is given in Section \ref{ch4:proofMainTheorem}.
Our approach to show existence of such solution is the following: first we split this problem
into two auxiliary ones and define a proper mapping, for which the solution to our problem
is a fixed point. Then we give suitable estimates to show that the mapping is a contraction.
The last step, together with the Banach fixed point theorem, gives us existence of the solution.

To obtain asymptotic behaviour of the fluid we consider the  Oseen system:
\begin{eqnarray}
	u_\infty \overline{v}_{,1} - \Delta \overline{v} - \nabla \overline{p} & = & F \qquad \textrm{in~} \mathbb{R}^2,\label{Oseen10}\\
	\mathrm{div~}\overline{v} & = & 0 \qquad \textrm{in~} \mathbb{R}^2,\label{Oseen20}\\
	\overline{v} & \to & 0 \qquad \textrm{as~}|x|\to\infty\label{Oseen30}.
\end{eqnarray}
Once one has result of Theorem \ref{ch4:mainTheorem} he can show existence of solution $(\overline{v}, \overline{p})$
to this system in the same class of functions, that is $\hat{\overline{v}}\in \mathcal{X}_\beta$ and $\hat{\overline{p}}\in \mathcal{Y}$.
We consider this system to show the following theorem describing asymptotic profile of solutions to the Navier-Stokes
equations and the Oseen system:
\begin{Theorem}\label{mainTheoremOseenProfile}
	Considering the solution $(v, p)$ to the system (\ref{ch4:intr30})-(\ref{ch4:intr50}) and
	the solution $(\overline{v}, \overline{p})$ to the system (\ref{Oseen10})-(\ref{Oseen30}), both
	with the same force $F$, $\beta$ and $u_\infty$. Then the following statement is true:
	\begin{equation}
	 	\hat v - \hat{\overline{v}} \in \mathcal{X}_{2\beta},
	\end{equation}
	that is the difference $\hat v - \hat{\overline{v}}$ has a better decay at infinity than
	$v$ and $\overline{v}$ themselves.
\end{Theorem}

The paper is organized as follows: in the next section we introduce two auxiliary problems and derive
formulas for their solutions by means of the Fourier transform. 
Then we provide suitable estimates, which play fundamental role in the proof of existence of a solution
to the main problem (\ref{ch4:intr30})-(\ref{ch4:intr50}). Section \ref{ch4:mainLemmas} is
devoted to technical lemmas, which play fundamental role in mentioned estimates. We choose this ``nonlocal'' form of presentation of proofs hoping it is more convinient for the Reader. For the same reason we include calculations of all cases even though sometimes they
present similarities.

\section{Auxiliary systems}

This section is devoted to a reformulation of the original Navier-Stokes equations. The Stokes system can not be recognized as  purely parabolic system, however we are able to represent the pressure in terms of the  velocity  by a nonlocal operator. Then our problem is reduced to two equations.

The first one is for the \textbf{pressure}.
Taking $\mathrm{div~}$ from (\ref{ch4:intr30}) we have:
\begin{equation}\label{ref1}
	\Delta p = \mathrm{div~}F - \mathrm{div~}\mathrm{div~}(v\otimes v),
\end{equation}
since $\mathrm{div~}v = 0$ and $v\cdot\nabla v = \mathrm{div~}(v\otimes v)$. We introduce function $G$
as the right hand side of (\ref{ref1}), i.e.
\begin{equation*}
	G = \mathrm{div~}F - \mathrm{div~}\mathrm{div~}(v\otimes v).
\end{equation*}

The second one is for the \textbf{velocity}. We transform (\ref{ch4:intr30}) into the following system:
\begin{eqnarray}\label{ref2}
	u_\infty v_{,1} - \Delta v & = & \nabla p - \mathrm{div~}(v\otimes v) \qquad \textrm{in~} \mathbb{R}^2,\label{ch4:intr80}\\
	v &\to& 0 \qquad \textrm{as~~~} |x|\to\infty.
\end{eqnarray}
As earlier -- we introduce function $H$ as the right hand side of (\ref{ch4:intr80}), i.e.
\begin{equation*}	
	H = F+\nabla p - \mathrm{div~}(v\otimes v).
\end{equation*}

Our mapping used in fixed point theorem will be considered as follows: we start with $v$ in a proper
Banach space, then we calculate pressure $p$ from (\ref{ref1}). Having $v$ and $p$ we may use
(\ref{ch4:intr80}) to calculate new $\tilde v$. We show that this mapping maps a suitable ball into itself
assuring that there exists $v$ for which $\tilde v = v$.

In this following sections we deal with our two problems (\ref{ref1}) and (\ref{ch4:intr80}) 
using the Fourier transform in $x_2$ space variable and transforming them into ordinary differential equations.
This point of view at the Stokes system via equations (\ref{ref1})-(\ref{ref2}) was effectively used in \cite{DaMu,DH} for the issue of existence in the maximal regularity regime.
Similar procedure was used in \cite{Wittwer1}, however not for the velocity directly, but for the rotation of the fluid. 

\subsection{Derivation of the solution}

Let us focus on \textbf{the system for the pressure} (\ref{ref1}) first. Taking Fourier transform in $x_2$ variable and denoting the new variable
as $\xi$ and $x_1$ as $t$ we get:
\begin{equation}\label{ch4:intr100}
	\ddot{\widehat{p}} - \xi^2\widehat{p} = \widehat{G}.
\end{equation}
For the simplicity we omit the hat $\widehat{}$.

Introducing $w = \dot p$ we can rewrite (\ref{ch4:intr100}) as:
\begin{equation}\label{ch4:intr110}
	\dot{\vecDD{p}{w}} = \left[ 
	\begin{array}{cc}
		0 & 1\\
		\xi^2 & 0
		\end{array}
	\right] {\vecDD{p}{w}} + {\vecDD{0}{G}}.
\end{equation}
Eigenvalues and eigenvectors can be easily computed: $\lambda_1 = -|\xi|$, $\lambda_2 = |\xi|$ and
$\varphi_1 = [-1/|\xi|, 1]$, $\varphi_2 = [1/|\xi|, 1]$. Introducing matrix $P = [\varphi_1, \varphi_2]$,
i.e.
\begin{equation*}
	P =  \left[ 
	\begin{array}{cc}
		-\frac{1}{|\xi|} & \frac{1}{|\xi|}\\
		1 & 1
		\end{array}
	\right],
\end{equation*}
and new variables $[U_1, U_2] = P^{-1}[p, w]$ we rewrite (\ref{ch4:intr110}) as:
\begin{equation*}
	\dot{\vecDD{U_1}{U_2}} = \left[ 
	\begin{array}{cc}
		-|\xi| & 0\\
		0 & |\xi|
		\end{array}
	\right] 
	\vecDD{U_1}{U_2} + P^{-1}\vecDD{0}{G}.
\end{equation*}
A solution to this system is:
\begin{eqnarray*}
	U_1(t, \xi) & = & \frac{1}{2}\int_{-\infty}^t e^{|\xi|(s-t)}G(s) ds, \\
	U_2(t, \xi) & = & -\frac{1}{2}\int_{t}^{\infty} e^{-|\xi|(s-t)} G(s) ds
\end{eqnarray*}
which gives us:
\begin{eqnarray}\label{ch4:intr140}
p(t, \xi) & = & -\frac{1}{2}\frac{1}{|\xi|}\left( \int_{-\infty}^{t} e^{-|\xi||t-s|} G(s, \xi) ds + \int_{t}^{\infty} e^{-|\xi||t-s|} G(s, \xi) ds,\right)	\\
	& = & -\frac{1}{2|\xi|}\int_{-\infty}^{+\infty} e^{-|\xi||t-s|} G(s, \xi) ds.
\end{eqnarray}

\bigskip
For \textbf{the system for the velocity} (\ref{ref2}) we may present a similar approach. 
Using Fourier transform in $x_2$ variable for the system (\ref{ref2}) and taking the right hand side to be $\delta$ we obtain equation for the fundamental solution:
\begin{equation}\label{fundV}
 u_\infty \dot{V}(t, \xi) - \ddot{V}(t, \xi) + \xi^2 V(t, \xi) = \delta.
\end{equation}
Now denoting
\begin{equation*}
\Delta = \sqrt{u_\infty^2+4\xi^2}, \quad\lambda_1 = \frac{1}{2}(u_\infty-\sqrt{u_\infty^2+4\xi^2}),\quad \lambda_2 = \frac{1}{2}(u_\infty+\sqrt{u_\infty^2+4\xi^2})
\end{equation*}
we may write the fundamental solution $V$ as follows:
\begin{equation}
 V(t, \xi) = \frac{1}{\Delta} 
	\begin{cases}
	 e^{\lambda_2 t} & \mathrm{for~} t < 0, \\
	 e^{\lambda_1 t} & \mathrm{for~} t > 0.
	\end{cases}
\end{equation}
Now if $v(t, \xi)$ is a solution to (\ref{ref2}) with $H(t, \xi)$ as the right hand side we get:
\begin{multline}\label{ch4:intr170}
 v(t, \xi) = (V\ast H)(t, \xi) = \int_{-\infty}^{t} V(t-s,\xi) H(s) ds + \int_{t}^\infty V(t-s, \xi) H(s) ds = \\
	-\frac{1}{\Delta} \int_{-\infty}^{t} e^{-\lambda_1(s-t)} H(s, \xi) ds + \frac{1}{\Delta}\int_{t}^\infty e^{-\lambda_2(s-t)}H(s, \xi) ds.
\end{multline}

\bigskip
Let us now focus on detailed information about $G(s)$ and $H(s)$. We start with the former. Since
\begin{equation*}
	G(s) = \mathcal{F}_{x_2}(\mathrm{div~}F - \mathrm{div~}\mathrm{div~}(v\otimes v))
\end{equation*}
we see that
\begin{equation*}
	G(s, \xi) = \partial_s \widehat{F_1}(s, \xi) + i\xi\widehat{F_2}(s, \xi) -
		\partial^2_s \widehat{(v_1^2)} + 2\partial_s i\xi \widehat{(v_1v_2)} - \xi^2 \widehat{(v_2^2)}.
\end{equation*}
First we integrate by parts the term from (\ref{ch4:intr140}) corresponding to $\partial^2_s\widehat{(v_1^2)}$, namely:
\begin{eqnarray*}
	\int_{-\infty}^\infty e^{-|\xi||t-s|} \partial^2_s \widehat{(v_1^2)} ds & = & 
		-\int_{-\infty}^t |\xi|e^{-|\xi|(t-s)} \partial_s \widehat{(v_1^2)} ds+ \partial_s \widehat{(v_1^2)}(t)\\
		&  & -\int_{t}^{\infty} (-|\xi|)e^{-|\xi|(s-t)}\partial_s\widehat{(v_1^2)} ds - \partial_s \widehat{(v_1^2)}(t)\\
		& = & \int_{-\infty}^t(|\xi|^2)e^{-|\xi|(t-s)} \widehat{(v_1^2)} ds- |\xi| \widehat{(v_1^2)}(t) \\
		&  & -\int_{t}^{\infty} (-|\xi|^2)e^{-|\xi|(s-t)} \widehat{(v_1^2)} ds - |\xi| \widehat{(v_1^2)}(t)\\
		& = & \int_{-\infty}^{\infty} |\xi|^2 e^{-|\xi||t-s|} \widehat{(v_1^2)} ds - 2|\xi| \widehat{(v_1^2)}(t),
\end{eqnarray*}
hence
\begin{equation}\label{ch4:intr200}
	\frac{1}{2|\xi|}\int_{-\infty}^\infty e^{-|\xi||t-s|} \partial^2_s \widehat{(v_1^2)} ds = 
			\frac{1}{2}\int_{-\infty}^{\infty} |\xi| e^{-|\xi||t-s|} \widehat{(v_1^2)} ds - \widehat{(v_1^2)}(t).
\end{equation}
In the same manner, terms from (\ref{ch4:intr140}) corresponding to $2 i\xi \partial_s \widehat{(v_1v_2)}$, $\xi^2 \widehat{(v_2^2)}$ are of the same
structure as the first term on the right hand side of (\ref{ch4:intr200}). Similarly $\partial_s \widehat{F_1}$ and $i\xi \widehat{F_2}$
can be considered as one term. 
Summarizing, $p$ can be presented as
\begin{multline}\label{ch4:intr210}
	p(t, \xi) = -\frac{1}{2}\int_{-\infty}^{\infty} |\xi| e^{-|\xi||t-s|} 
			\left(\sum_{ij}c_{ij}\widehat{(v_i v_j)}\right) ds\\
			-\frac{1}{2}\int_{-\infty}^{\infty} e^{-|\xi||t-s|} \left(\sum_i b_i \widehat{F_i}\right) ds
			+ \widehat{(v_1)^2}(t),
\end{multline}
for some constants $c_{ij}$ and $b_i$ such that $|c_{ij}| = |b_i| = 1$ (which is irrelevant for our purposes).

The same calculations can be repeated for $v$ and $H$, i.e. for (\ref{ch4:intr170}) to get that
\begin{multline}\label{ch4:intr220}
	v_1(t, \xi) = \frac{\lambda_1}{\Delta}\int_{-\infty}^t e^{-\lambda_1(s-t)} \widehat{(p(\xi, s)-v_1^2)}ds - \frac{\lambda_2}{\Delta}\int_t^{\infty} e^{-\lambda_2(s-t)} \widehat{(p(\xi, s)+v_1^2)} ds \\
		 + \frac{i\xi}{\Delta}\int_{-\infty}^t e^{-\lambda_1(s-t)} (\widehat{(v_1v_2)}-\frac{\hat{F}_1}{i\xi}) ds + \frac{i\xi}{\Delta}\int_{t}^\infty e^{-\lambda_2(s-t)} (\widehat{(v_1v_2)}+\frac{\hat{F}_1}{i\xi}) ds,
\end{multline}
and
\begin{multline*}
	v_2(t, \xi)  = \frac{\lambda_1}{\Delta}\int_{-\infty}^t e^{-\lambda_1(s-t)} \widehat{(v_1v_2)}ds - \frac{\lambda_2}{\Delta}\int_t^{\infty} e^{-\lambda_2(s-t)} \widehat{(v_1v_2)} ds \\
		 + \frac{i\xi}{\Delta}\int_{-\infty}^t e^{-\lambda_1(s-t)} (\widehat{(p - v_2^2)}-\frac{\hat{F}_2}{i\xi}) ds + \frac{i\xi}{\Delta}\int_{t}^\infty e^{-\lambda_2(s-t)} (\widehat{(p-v_2^2)}+\frac{\hat{F}_2}{i\xi}) ds.
\end{multline*}

\subsection{Main estimates.}

We are now in position to formulate Lemmas which play fundamental role in showing that our mapping is a contraction.

\begin{Lemma}\label{ch4:pEstimate}
	Let $\widehat{v}\in \mathcal{X}_\beta$, such that $\|\widehat{v}\|_{\mathcal{X}_\beta} \leq M$,
	and $\widehat{F}/\xi\in \mathcal{Y}$, such that $\|\widehat{F}/\xi\|_{\mathcal{Y}} \leq N_F$.
	Given $p$ in the form:
	\begin{multline}\label{ch4:pForm}
		\widehat{p}(t, \xi) = -\frac{1}{2}\int_{-\infty}^{\infty} |\xi| e^{-|\xi||t-s|} 
				\left(\sum_{ij}c_{ij}\widehat{(v_i v_j)}\right) ds\\
				-\frac{1}{2}\int_{-\infty}^{\infty} e^{-|\xi||t-s|} \left(\sum_i b_i \widehat{F_i}\right) ds
				+ \widehat{(v_1)^2}(t).
	\end{multline}
	Then $\widehat{p}\in \mathcal{Y}$ and
	\begin{equation*}
		\|\widehat{p}\|_{\mathcal{Y}} \leq Cu_\infty^{-2\beta+1/2}M^2 + N_F,
	\end{equation*}
	for some constant $C$ independent of $u_\infty$ and $F$.
\end{Lemma}

\begin{Proof}{}
	To prove this Lemma we use results from Section \ref{ch4:mainLemmas}, that is:
	recalling that $|c_{ij}| = |b_i| = 1$ we estimate $\widehat{p}$ as follows:
	\begin{multline}\label{ch4:est10}
		|\widehat{p}(t, \xi)| \leq \frac{1}{2}\int_{-\infty}^{\infty} |\xi| e^{-|\xi||t-s|} 
				\left(\sum_{ij}|\widehat{(v_i v_j)|}\right) ds\\
				+\frac{1}{2}\int_{-\infty}^{\infty} e^{-|\xi||t-s|} \left(\sum_i |\widehat{F_i}|\right) ds
				+ \widehat{|(v_1)^2|}(t).
	\end{multline}
	Since $\widehat{v} \in \mathcal{X}_\beta$ we use estimate (\ref{lem50est2}) form Lemma \ref{new:mainLem50} to get that:
	\begin{equation*}
		\|\widehat{(v_iv_j)}\|_{\mathcal{Y}} = \|\widehat{v_i}\ast \widehat{v_j}\|_{\mathcal{Y}} \leq Cu_\infty^{-2\beta+1/2} M^2.
	\end{equation*}
	It estimates the last term on the right hand side of (\ref{ch4:est10}), but also, with a help of estimate (\ref{lem110est2}) from Lemma \ref{new:mainLem110}, gives us
	the following estimate:
	\begin{equation*}
		\left\|\frac{1}{2}\int_{-\infty}^{\infty} |\xi| e^{-|\xi||t-s|} 
				\left(\sum_{ij}|\widehat{(v_i v_j)|}\right) ds \right\|_{\mathcal{Y}} \leq 4Cu_\infty^{-2\beta+1/2}M^2.
	\end{equation*}
	To finish estimate of $\widehat{p}$ we present the remaining term and estimate it as follows
	\begin{multline*}
		\left\|\frac{1}{2}\int_{-\infty}^{\infty} e^{-|\xi||t-s|} \left(\sum_i |\widehat{F_i}|\right) ds \right\|_{\mathcal{Y}}\leq\\
		\left\|\frac{1}{2}\int_{-\infty}^{\infty} |\xi|e^{-|\xi||t-s|} \left(\sum_i |\widehat{F_i}|/|\xi|\right) ds\right\|_{\mathcal{Y}}
		\leq \|\widehat{F_i}/|\xi|\|_{\mathcal{Y}} = N_F,
	\end{multline*}
	where again we used (\ref{lem110est2}).
\end{Proof}

The second Lemma we need to prove is the following:
\begin{Lemma}\label{ch4:vEstimate}
	Let $\widehat{w}\in \mathcal{X}_\beta$, such that $\|\widehat{w}\|_{\mathcal{X}_\beta} \leq M$,
	and $\widehat{p}\in \mathcal{Y}$, such that $\|\widehat{p}\|_{\mathcal{Y}} \leq N$.
	Then, for the following terms $\hat v_1$ and $\hat v_2$:
	\begin{multline}\label{ch4:est60}
		\hat v_1(t, \xi) = \frac{\lambda_1}{\Delta}\int_{-\infty}^t e^{-\lambda_1(s-t)} \widehat{(p(\xi, s)-w_1^2)}ds - \frac{\lambda_2}{\Delta}\int_t^{\infty} e^{-\lambda_2(s-t)} \widehat{(p(\xi, s)+w_1^2)} ds \\
			 + \frac{i\xi}{\Delta}\int_{-\infty}^t e^{-\lambda_1(s-t)} (\widehat{(w_1w_2)}-\frac{\hat{F}_1}{i\xi}) ds + \frac{i\xi}{\Delta}\int_{-\infty}^t e^{-\lambda_2(s-t)} (\widehat{(w_1w_2)} +\frac{\hat{F}_1}{i\xi})ds,
	\end{multline}
	and
	\begin{multline}\label{ch4:est70}
		\hat v_2(t, \xi)  = \frac{\lambda_1}{\Delta}\int_{-\infty}^t e^{-\lambda_1(s-t)} \widehat{(w_1w_2)}ds - \frac{\lambda_2}{\Delta}\int_t^{\infty} e^{-\lambda_2(s-t)} \widehat{(w_1w_2)} ds \\
			 + \frac{i\xi}{\Delta}\int_{-\infty}^t e^{-\lambda_1(s-t)} (\widehat{(p - w_2^2)}-\frac{\hat{F}_2}{i\xi}) ds + \frac{i\xi}{\Delta}\int_{-\infty}^t e^{-\lambda_2(s-t)} (\widehat{(p-w_2^2)}+\frac{\hat{F}_2}{i\xi}) ds.
	\end{multline}
	the following estimate is valid:
	\begin{equation}
		\|\hat v_i\|_{\mathcal{X}_\beta} \leq C(\|\hat p\|_{\mathcal{Y}} + u_\infty^{-2\beta+1/2} \|\hat w\|_{\mathcal{X}_\beta}^2 + \|\hat F/\xi\|_{\mathcal{Y}}) 
			= C(N + u_\infty^{-2\beta+1/2} M^2 + \|\hat F/\xi\|_{\mathcal{Y}}).
	\end{equation}
\end{Lemma}

\begin{Proof}{}
	The proof is analogous to the proof of the previous lemma. First we notice, that
	\begin{equation*}
		\|\widehat{(w_iw_j)}\|_{\mathcal{Y}} = \|\widehat{w_i}\ast \widehat{w_j}\|_{\mathcal{Y}} \leq Cu_\infty^{-2\beta+1/2} M^2,
	\end{equation*}
	like also $\|p\|_{\mathcal{Y}} \leq N$.
	
	Then, we find, that in the form of $v_i$, i.e. in (\ref{ch4:est60}) and (\ref{ch4:est70}), these integrals in a sequence
	are respectively in the form of integrals $\tilde{B}$, $\tilde{D}$, $\tilde{A}$, $\tilde{C}$ from Lemma \ref{ch4:mainLem195}.
	Thus, applying this Lemma we obtain:
		\begin{equation*}
			\|v_i\|_{\mathcal{X}_\beta} \leq C(\|p\|_{\mathcal{Y}} + \|\widehat{(w_iw_j)}\|_{\mathcal{Y}}) \leq C(N + u_\infty^{-2\beta+1/2} M^2).
		\end{equation*}
\end{Proof}

\section{Main results.}

In this section we gather our main result on existence of the solution to our problem. 

\subsection{The proof of Theorem \ref{ch4:mainTheorem}.}\label{ch4:proofMainTheorem}

% label_prefix:ex

In this section we would like to prove existence of a solution to our problem. We use standard approach, namely
Banach's fixed point theorem for a contraction mapping.

We recall, that our mapping is defined as follows: having a vector field $w$ and force $F$ we calculate the pressure
$p$ using formula (\ref{ch4:pForm}), then we calculate a vector field $v$ using $p$, $w$ and formula (\ref{ch4:est60})-(\ref{ch4:est70}).
Let us denote this mapping as $G : \mathcal{X}_\beta \to \mathcal{X}_\beta$. First we would like to show, that 
there exists a constant $\epsilon$ such that for sufficiently small $\|F/|\xi|\|_{\mathcal{Y}}$ the mapping $G$
maps a ball of radius $\epsilon$ in the space $\mathcal{X}_\beta$ into itself, namely:
\begin{equation}\label{ch4:ex10}
	G(\mathcal{B}_\alpha(\epsilon) ) \subset \mathcal{B}_\alpha(\epsilon).
\end{equation}

We take $w \in \mathcal{X}_\beta$ such that $\|w\|_{\mathcal{X}_\beta} \leq M$ and $F$ such that
$\|\hat F/|\xi|\|_{\mathcal{Y}} \leq N_F$. From Lemma \ref{ch4:pEstimate} we have:
\begin{equation*}
	\|p\|_{\mathcal{Y}} \leq C u_\infty^{-2\beta + 1/2} M^2 + N_F.
\end{equation*}
We may now use Lemma \ref{ch4:vEstimate} to obtain:
\begin{equation*}
	\|v\|_{\mathcal{X}_\beta} \leq C(\|p\|_{\mathcal{Y}}+ \|\hat F/\xi\|_{\mathcal{Y}} + u_\infty^{-2\beta + 1/2} M^2) \leq
		C(u_\infty^{-2\beta+1/2} M^2 + N_F).
\end{equation*}
To find $\epsilon$ in (\ref{ch4:ex10}) we have to solve an inequality:
\begin{equation*}
	C(u_\infty^{-2\beta+1/2} M^2 + N_F) \leq M.
\end{equation*}
From this inequality we can read the following constraints:
\begin{equation}
 N_F \leq \frac{1}{4C^2}u_\infty^{2\beta - 1/2},\qquad \epsilon = \frac{1}{2C}u_\infty^{2\beta-1/2}.
\end{equation}
For this choice of $N_F$ and $\epsilon$ one gets $G$ to map a ball $B_\alpha(\epsilon)$ into itself.

In a similar way we show, that on a smaller ball the mapping $G$ is a contraction. For $w_1, w_2 \in \mathcal{B}_\beta(\epsilon/2)$ 
and corresponding $v_1$, $v_2$ we have:
\begin{equation*}
	\|v_1 - v_2\|_{\mathcal{X}_\beta} \leq C u_\infty^{-2\beta + 1/2} \|w_1 - w_2 \|_{\mathcal{X}_\beta}^2 \leq
		\gamma \|w_1 - w_2\|_{\mathcal{X}_\beta},
\end{equation*}
where $\gamma < 1$. Using the Banach fixed point theorem we get, that there exists a vector field $v \in \mathcal{B}_\beta(\epsilon/2)$,
such that $v = G(v)$.

\subsection{The proof of the Theorem \ref{mainTheoremOseenProfile}.}

In order to investigate the asymptotic profile of solutions to the Navier-Stokes equations we observe
that the same existence argument as was presented in the preceeding secion can be adapted fot the Oseen system to obtain:
Obviously one can repeat the same existence argument for the Oseen system (\ref{Oseen10})-(\ref{Oseen30}) to get:
the solution $(\overline{v}, \overline{p})$ such that $\overline{v}\in\mathcal{X}_\beta$ and $\overline{p}\in\mathcal{Y}$.

This solution satisfies the following identities (we focus only on $\overline{v}_1$ and $\overline{p}$):
\begin{multline}
	\overline{v}_1(t, \xi) = \frac{\lambda_1}{\Delta}\int_{-\infty}^t e^{-\lambda_1(s-t)} \widehat{\overline{p}(\xi, s)}ds - \frac{\lambda_2}{\Delta}\int_t^{\infty} e^{-\lambda_2(s-t)} \widehat{\overline{p}(\xi, s)} ds \\
		 + \frac{i\xi}{\Delta}\int_{-\infty}^t e^{-\lambda_1(s-t)} (-\frac{\hat{F}_1}{i\xi}) ds + \frac{i\xi}{\Delta}\int_{t}^\infty e^{-\lambda_2(s-t)} \frac{\hat{F}_1}{i\xi} ds,
\end{multline}
\begin{equation}
	\widehat{\overline{p}}(t, \xi) = -\frac{1}{2}\int_{-\infty}^{\infty} e^{-|\xi||t-s|} \left(\sum_i b_i \widehat{F_i}\right) ds.
\end{equation}
For the readability of the paper we recall identities for the solution $(v, p)$ of the Navier-Stokes equations:
\begin{multline}
	v_1(t, \xi) = \frac{\lambda_1}{\Delta}\int_{-\infty}^t e^{-\lambda_1(s-t)} \widehat{(p(\xi, s)-v_1^2)}ds - \frac{\lambda_2}{\Delta}\int_t^{\infty} e^{-\lambda_2(s-t)} \widehat{(p(\xi, s)+v_1^2)} ds \\
		 + \frac{i\xi}{\Delta}\int_{-\infty}^t e^{-\lambda_1(s-t)} (\widehat{(v_1v_2)}-\frac{\hat{F}_1}{i\xi}) ds + \frac{i\xi}{\Delta}\int_{t}^\infty e^{-\lambda_2(s-t)} (\widehat{(v_1v_2)}+\frac{\hat{F}_1}{i\xi}) ds,
\end{multline}
\begin{multline}
	\widehat{p}(t, \xi) = -\frac{1}{2}\int_{-\infty}^{\infty} |\xi| e^{-|\xi||t-s|} 
			\left(\sum_{ij}c_{ij}\widehat{(v_i v_j)}\right) ds\\
			-\frac{1}{2}\int_{-\infty}^{\infty} e^{-|\xi||t-s|} \left(\sum_i b_i \widehat{F_i}\right) ds
			+ \widehat{(v_1)^2}(t).
\end{multline}
We are interested in an estimate of the difference $v_1 - \overline{v}_{1}$. For that we first find:
\begin{equation}\label{new:pressureDifference}
	p(t, \xi) - \overline{p}(t, \xi) = -\frac{1}{2}\int_{-\infty}^{\infty} |\xi| e^{-|\xi||t-s|} \left(\sum_{ij}c_{ij}\widehat{(v_i v_j)}\right) ds + \widehat{(v_1)^2}(t).
\end{equation}
Using our assumptions on $v$ and using Lemma \ref{new:mainLem50} we see that all convolutions appearing in (\ref{new:pressureDifference}) can be estimated
in a pointwise manner by the term $|t|^{-1/2}(u_\infty+t\xi^2)^{-2\beta+1/2}$. Then one may use Lemma \ref{new:mainLem110} to get the following pointwise estimate
for the difference of pressures:
\begin{equation}
	|p(t, \xi) - \overline{p}(t, \xi)| \leq C |t|^{-1/2}(u_\infty + |t\xi^2|)^{-2\beta + 1/2}.
\end{equation}

To estimate the difference of velocities first we denote 
$\pi(t, \xi) = p(t, \xi) - \overline{p}(t, \xi)$ and write this difference as follows:
\begin{multline}
	v_1(t, \xi) - \overline{v_1}(t, \xi) =\\
	= \frac{\lambda_1}{\Delta}\int_{-\infty}^t e^{-\lambda_1(s-t)} \widehat{(\pi(\xi, s)-v_1^2)}ds - \frac{\lambda_2}{\Delta}\int_t^{\infty} e^{-\lambda_2(s-t)} \widehat{(\pi(\xi, s)+v_1^2)} ds \\
		 + \frac{i\xi}{\Delta}\int_{-\infty}^t e^{-\lambda_1(s-t)} \widehat{(v_1v_2)} ds + \frac{i\xi}{\Delta}\int_{t}^\infty e^{-\lambda_2(s-t)} \widehat{(v_1v_2)} ds.
\end{multline}
Again, using results of Lemma \ref{new:mainLem50} to estimate convolutions in above equations
we combine them with the previous estimate of $\pi(t, \xi)$ and we write:
\begin{equation}
	|\widehat{\pi}(t, \xi)| + |\widehat{v_1^2}(t, \xi)| + |\widehat{v_1v_2}(t, \xi)| \leq
		C |t|^{-1/2}(u_\infty + |t\xi|^2)^{-2\beta + \frac{1}{2}}.
\end{equation}

Now we are in position to use Lemma \ref{new:mainLem195} and complete the proof of Theorem \ref{mainTheoremOseenProfile}.

\section{Main Lemmas}\label{ch4:mainLemmas}

The main auxiliary lemma, which will be used many times is the following:
\begin{Lemma}
	Given $\theta > 0$. Then the following estimate is valid:
	\begin{equation}\label{ch4:mainIneq}
		\int_{0}^\infty e^{-u} |u-\theta|^{-1/2} du \leq (1+\theta)^{-1/2}.
	\end{equation}
\end{Lemma}

\begin{Proof}{}
	Let $I = \int_{0}^\infty e^{-u} |u-\theta|^{-1/2} du$. We split this integral into three parts:
	\begin{equation*}
		I = I_1 + I_2 + I_3	= \int_{0}^{\theta/2} + \int_{\theta/2}^{2\theta} + \int_{2\theta}^\infty,
	\end{equation*}
	and estimate them separately. Let us focus on $I_1$:
	\begin{equation}
		I_1 = \int_{0}^{\theta/2}e^{-u} |u-\theta|^{-1/2} du 
			\leq 2\int_{0}^{\theta/2} e^{-u} |\theta|^{-1/2}du = 2|\theta|^{-1/2}(1-e^{-\theta})
			\leq (1+\theta)^{-1/2},
	\end{equation}
	since $(1-e^{-\theta})\sim \theta/(1+\theta)$.
	
	For $I_2$ we proceed as follows:
	\begin{equation*}
		I_2 = \int_{\theta/2}^{2\theta}e^{-u} |u-\theta|^{-1/2} du \leq C\int_{0}^{2\theta}e^{-\theta} |u|^{-1/2} du 
			\leq Ce^{-\theta}|\theta|^{1/2} \leq \frac{C}{(1+\theta)^{1/2}},
	\end{equation*}
	which is a desired estimate.
	
	Integral $I_3$ we estimate for small and large $\theta$ ($\theta < 1$ and $\theta > 1$ respectively). In the first case
	\begin{equation*}
		I_3 = \int_{\theta}^1 + \int_1^\infty = (1-\theta^{1/2}) + C \leq C \leq (1+\theta)^{-1/2}
	\end{equation*}
	while in the second one we have:
	\begin{equation*}
		I_3 \leq \int_{\theta}^{\infty} e^{-u} u^{-1/2} \leq \int_{\theta}^\infty e^{-u} |\theta|^{-1/2} \leq (1+\theta)^{-1/2}.
	\end{equation*}
\end{Proof}

We also consider the following modification of the previous Lemma:
\begin{Lemma}
	Given $\theta > 0$. Then the following estimate is valid:
	\begin{equation}\label{new:mainIneq}
		\int_{0}^\infty e^{-u} |u-\theta|^{-1/2} (1 + |u-\theta|)^{-2\beta+1/2} du \leq C(1+|\theta|)^{-2\beta}.
	\end{equation}
\end{Lemma}

\begin{Proof}{}
	Let $I = \int_{0}^\infty e^{-u} |u-\theta|^{-1/2} (1 + |u-\theta|)^{-2\beta+1/2} du$. We split this integral into three parts:
	\begin{equation*}
		I = I_1 + I_2 + I_3	= \int_{0}^{\theta/2} + \int_{\theta/2}^{2\theta} + \int_{2\theta}^\infty,
	\end{equation*}
	and estimate them separately. Let us focus on $I_1$:
	\begin{multline*}
		I_1 = \int_{0}^{\theta/2}e^{-u} |u-\theta|^{-1/2} (1 + |u-\theta|)^{-2\beta + 1/2} du 
			 \leq 2\int_{0}^{\theta/2} e^{-u} |\theta|^{-1/2} (1 + |\theta|)^{-2\beta + 1/2} du \\
		    = 2|\theta|^{-1/2} (1 + |\theta|)^{-2\beta + 1/2}(1-e^{-\theta}) \leq (1+|\theta|)^{-2\beta},
	\end{multline*}
	since $(1-e^{-\theta}) \sim \theta/(1+\theta)$.

	For $I_2$ we proceed as follows:
	\begin{equation*}
		I_2 = \int_{\theta/2}^{2\theta}e^{-u} |u-\theta|^{-1/2} (1 + |u-\theta|)^{-2\beta + 1/2} du 
		    \leq Ce^{-\theta} \int_{0}^{2\theta} |u|^{-1/2}(1 + |u|)^{-2\beta + 1/2}  du =: \widetilde{I_2}.
	\end{equation*}
	Now we distinguish two cases: first $\theta \leq 1$. Then we have:
	\begin{equation}
		\widetilde{I_2} \leq C e^{-\theta}|\theta|^{1/2} \leq C (1+|\theta|)^{-2\beta}. 
	\end{equation}
	In the second case $\theta > 1$ we simply have:
	\begin{equation}
		\widetilde{I_2} \leq e^{-\theta}\int_0^\theta |u|^{-1/2}(1+|u|)^{-2\beta+1/2} du \leq C(\beta)p(\theta) e^{-\theta} \leq C(\beta)(1+\theta)^{-2\beta},
	\end{equation}
	where $p(\theta)$ is a suitable polynomial with respect to $\theta$.
	Integral $I_3$ we first rewrite as $I_3 = e^{-\theta}\int_\theta^\infty e^{-u}|u|^{-1/2}(1+|u|)^{-2\beta+1/2} du$.
	Then since $\beta > 1/4$ we have:
	\begin{equation}
		I_3 \leq e^{-\theta} (1 + \theta)^{-2\beta+1/2} \int_\theta^\infty e^{-u}|u|^{-1/2} \leq C (1 + \theta)^{-2\beta},
	\end{equation}
	which is the desired estimate.
\end{Proof}

% label_prefix:mainLem

\begin{Lemma}\label{new:mainLem50}
	Let $a, b \in \mathcal{X}_\beta$. Then for $\beta < 1/2$ the following estimate is valid
	\begin{equation}\label{lem50est1}
		(a\ast b)(t, \xi) \leq C \|a\|_{\mathcal{X}_\beta}\|b\|_{\mathcal{X}_\beta}|t|^{-1/2}(u_\infty + |t\xi^2|)^{-2\beta + \frac{1}{2}}.
	\end{equation}
	In particular for $\beta > 1/4$ one easily gets a weaker estimate:
	\begin{equation}\label{lem50est2}
	 	\|a\ast b\|_{\mathcal{Y}} \leq u_\infty^{-2\beta + 1/2}\|a\|_{\mathcal{X}_\beta}\|b\|_{\mathcal{X}_\beta}.
	\end{equation}
\end{Lemma}
\begin{Proof}{}
	Without any loss we may assume that $\|a\|_{\mathcal{X}_\beta} = \|b\|_{\mathcal{X}_\beta} = 1$.
	Therefore
	\begin{equation*}
		|a(t, \xi)| \leq (u_\infty+|t\xi^2|)^{-\beta}\qquad \textrm{for all~}(t, \xi)\in \mathbb{R}^2,
	\end{equation*}
	and we may write:
	\begin{equation}\label{new:mainLem60}
		|(a \ast b)(t, \xi)| \leq \int_{\mathbb{R}} \frac{1}{(u_\infty+|t||y-\xi|^2)^\beta}\frac{1}{(u_\infty+|ty^2|)^\beta} dy.
	\end{equation}
	By $I$ we denote the right hand side of (\ref{new:mainLem60}). Using the substitution $u = |t|^{1/2}y$ we have:
	\begin{equation}\label{new:mainLem70}
		I = |t|^{-1/2}\int_{\mathbb{R}} \frac{1}{(u_\infty+\left|u - |\xi||t|^{1/2}\right|^2)^\beta(u_\infty+|u|^2)^\beta}du.
	\end{equation}
	Because of the presence of the term $|t|^{-1/2}$ it is sufficient to estimate the integral in (\ref{new:mainLem70}).
	
	We split domain $\mathbb{R}$ in the integral into three parts:
	\begin{itemize}
		\item $A_1 = \{u : |u|\leq \frac{1}{2}|t|^{1/2}|\xi| \},$
		\item $A_2 = \{u : \frac{1}{2}|t|^{1/2}|\xi| < |u| \leq 2|t|^{1/2}|\xi|\},$
		\item $A_3 = \{u : 2|t|^{1/2}|\xi| < |u| \},$
	\end{itemize}
	and we denote by $J_1, J_2, J_3$ corresponding integrals. First we estimate $J_1$:
	\begin{multline*}
		J_1 = \int_{A_1} \frac{1}{(u_\infty+\left|u - |\xi||t|^{1/2}\right|^2)^\beta(u_\infty+|u|^2)^\beta}du \\
			\leq \int_{A_1} \frac{1}{(u_\infty+|\xi|^2|t|/4)^\beta(u_\infty+|u|^2)^\beta}du 
			\leq (u_\infty+|t||\xi|^2)^{-\beta} \int_{A_1} \frac{1}{(\sqrt{u_\infty}+|u|)^{2\beta}}du\\
			= \left.(u_\infty+|t||\xi|^2)^{-\beta}(\sqrt{u_\infty}+|u|)^{-2\beta+1}\right|_{0}^{|t|^{1/2}|\xi|/2}.
	\end{multline*}
	We use the assumption that $\beta < 1/2$ to get an estimate:
	\begin{equation*}
		J_1 \leq (u_\infty + |t\xi|^2)^{-2\beta+1/2}.
	\end{equation*}
	For $J_2$ we proceed in a similar way:
	\begin{multline*}
		J_2 = \int_{A_2} \frac{1}{(u_\infty+\left|u - |\xi||t|^{1/2}\right|^2)^\beta(u_\infty+|u|^2)^\beta}du \leq\\
			C\int_{A_2} \frac{1}{(u_\infty+\left|u - |\xi||t|^{1/2}\right|^2)^\beta(\sqrt{u_\infty}+|t|^{1/2}|\xi|)^{2\beta}}du,
	\end{multline*}
	using a substitution $t^{1/2}\xi - u = y$ we can write:
	\begin{equation*}
		J_2 \leq 2\int_{-t^{1/2}|\xi|}^{\frac{1}{2}t^{1/2}|\xi|}
			\frac{1}{(\sqrt{u_\infty}+|y|)^{2\beta}(\sqrt{u_\infty}+|t|^{1/2}|\xi|)^{2\beta}}dy
			\leq C(\sqrt{u_\infty}+|t|^{1/2}|\xi|)^{-4\beta + 1},
	\end{equation*}
	which is the same estimate as for $J_1$. Finally we estimate $J_3$:
	\begin{multline*}
		J_3 = \int_{A_3} \frac{1}{(u_\infty+\left|u - |\xi||t|^{1/2}\right|^2)^\beta(u_\infty+|u|^2)^\beta}du
			\\ \leq \int_{|t|^{1/2}|\xi|}^{\infty}\frac{1}{(\sqrt{u_\infty}+|u|)^{4\beta}} du
			= (\sqrt{u_\infty}+|t|^{1/2}|\xi|)^{-4\beta + 1}.
	\end{multline*}
	\noindent
	Gathering estimates for $J_1$, $J_2$ and $J_3$ we 
	the proof of Lemma \ref{new:mainLem50} is finished.
\end{Proof}

In order to estimate the difference of pressures we also need the following lemma:
\begin{Lemma}\label{new:mainLem110}
	Let $f$ satisfy the following inequality:
	\begin{equation}
		|f(t, \xi)|  \leq |t|^{-1/2}(u_\infty + |t\xi|^2)^{-2\beta + \frac{1}{2}}.
	\end{equation}
	Then for $1/2 > \beta > 1/4$ and $\tilde{I}$ defined as follows
	\begin{equation*}
		\tilde{I} := \xi\int_{\mathbb{R}} e^{-|\xi||t-y|} f(y, \xi) dy
	\end{equation*}
	the following estimate holds:
	\begin{equation}\label{lem110est1}
		|\tilde{I}(t, \xi)| \leq C |t|^{-1/2}(u_\infty + |t\xi|^2)^{-2\beta + \frac{1}{2}},
	\end{equation}
	where $C$ does not depend on $f$. 

	Moreover if $f\in \mathcal{Y}$ then the following estimate is valid
	\begin{equation}\label{lem110est2}
		\|\tilde{I}(t, \xi)\|_{\mathcal{Y}} \leq C u_\infty^{-2\beta + \frac{1}{2}}\|f\|_{\mathcal{Y}}.
	\end{equation}
\end{Lemma}
\begin{Proof}{}
	It is straightforward that we may focus on estimate of the following integral:
	\begin{equation*}
		I = |\xi|\int_{\mathbb{R}} e^{-|\xi||t-y|}  |y|^{-1/2}(u_\infty + |y\xi|^2)^{-2\beta + \frac{1}{2}} dy.
	\end{equation*}
	First we use a substitution $u = \xi y$ to get:
	\begin{equation}\label{new:mainLem140}
		I = |\xi|^{1/2}\int_{\mathbb{R}}e^{-||\xi t| -|u||} u^{-1/2} (u_\infty + |u\xi|)^{-2\beta+1/2} du.
	\end{equation}
	We split this integral into three parts: $A_1 = \{u : |u| < |\xi||t|/2 \}$,
	$A_2 = \{u : |\xi||t|/2 \leq u \leq 2|\xi||t|\}$, 
	$A_3 = \{ u : 2|\xi||t| \leq u\}$ and we introduce $I_1$, $I_2$ and $I_3$ as the 
	corresponding parts of integral $I$ from (\ref{new:mainLem140}).
	
	\noindent
	In what follows we denote $-2\beta + 1/2 = -\gamma$. Then condition $\beta < 1/2$ corresponds to $\gamma < 1/2$,
	while $\beta > 1/4$ implies $\gamma > 0$. 
	For $I_1$ we distinguish two cases: first for $|t\xi^2| < u_\infty$, and then for $|t\xi^2| \geq u_\infty$.
	In the former we have:
	\begin{multline*}
		I_1 	\leq \int_{0}^{|t\xi|/2} e^{-|t\xi|/2} |u|^{-1/2}|\xi|^{1/2}(u_\infty + u|\xi|)^{-\gamma} du 
			\leq  \int_{0}^{|t\xi|/2} e^{-|t\xi|/2} |u|^{-1/2}|\xi|^{1/2}(u_\infty)^{-\gamma} du \\
			 =    |t|^{-1/2}(u_\infty)^{-\gamma} e^{-|t\xi|/2}|t\xi|^1 
			\leq  |t|^{-1/2}(u_\infty + |t\xi^2|)^{-\gamma}.
	\end{multline*}
	which is the desired estimate.

	In the later case, that is for $|t\xi^2| \geq u_\infty$ one estimates integral $I_1$ as follows:
	\begin{multline*}
		I_1 	\leq  \int_{0}^{|t\xi|/2} e^{-|t\xi|/2} |u|^{-1/2}|\xi|^{1/2}(u_\infty + u|\xi|)^{-\gamma} du 
			\leq  \int_{0}^{|t\xi|/2} e^{-|t\xi|/2} |u|^{-1/2}|\xi|^{1/2}(u|\xi|)^{-\gamma} du \\
			 =    e^{-|t\xi|/2} |t\xi|^{-1/2-\gamma+1}|\xi|^{1/2-\gamma} 
			 =    e^{-|t\xi|/2} |t\xi^2|^{-\gamma} |t\xi|^{1} |t|^{-1/2} 
			\leq  |t|^{-1/2}(u_\infty+|t\xi^2|)^{-\gamma},
	\end{multline*}
	where we used the assumption that $\gamma < 1/2$. This completes estimate for $I_1$ for the first part of the lemma.

	For integral $I_2$ we proceed as follows.
	\begin{multline*}
	 I_2\leq |\xi|^{1/2}\int_{|\xi t|/2}^{2|\xi t|} e^{-||\xi t| - u|} u^{-1/2} (u_\infty+|u\xi|)^{-2\beta+1/2}du \\
	 	\leq |\xi|^{1/2} |\xi t| e^{-|\xi t|} |\xi t|^{-1/2}(u_\infty+|t\xi^2|)^{-2\beta+1/2}\\
	 	\leq |t|^{-1/2}\frac{|\xi t|}{(1+|\xi t|)}(u_\infty+|t\xi^2|)^{-2\beta+1/2}.
	\end{multline*}
	For $I_3$ we must distinguish two cases: one if $|\xi t| < 1$ and the other one for $|\xi t| \geq 1$. For the
	first case we have:
	\begin{eqnarray*}
		I_3  & \leq & (u_\infty+|t\xi^2|)^{-2\beta+1/2}|\xi|^{1/2}\int_{|\xi t|}^{\infty} e^{-u} u^{-1/2} du = (u_\infty+|t\xi^2|)^{-2\beta+1/2}(\int_{|\xi t|}^1 + \int_{1}^{\infty})\\
			& \leq & (u_\infty+|t\xi^2|)^{-2\beta+1/2}(|\xi|^{1/2}\int_{|\xi t|}^1 u^{-1/2}du + |\xi|^{1/2}\int_{1}^\infty e^{-u} u^{-1/2} du) \\
			& \leq & (u_\infty+|t\xi^2|)^{-2\beta+1/2}|\xi|^{1/2}((1-|\xi t|) + |\xi|^{1/2})\\
			& \leq & (u_\infty+|t\xi^2|)^{-2\beta+1/2}|\xi|^{1/2} \leq (u_\infty+|t\xi^2|)^{-2\beta+1/2}|t|^{-1/2},
	\end{eqnarray*}
	since $|\xi t| < 1$. For $|\xi t| \geq 1$ we have:
	\begin{equation*}
		I_3 \leq (u_\infty+|t\xi^2|)^{-2\beta+1/2}|\xi|^{1/2}\int_{|\xi t|}^\infty e^{-u} du = (u_\infty+|t\xi^2|)^{-2\beta+1/2}|\xi|^{1/2}e^{-|\xi t|},
	\end{equation*}
	and again using (\ref{ch4:mainLem150}) we get
	\begin{equation*}
		I_3 \leq C|t|^{-1/2}(u_\infty+|t\xi^2|)^{-2\beta+1/2}.
	\end{equation*}
	This completes the first part of the Lemma. 

	To show (\ref{lem110est2}) we proceed in a similar way. Since $f\in \mathcal{Y}$ we have
	\begin{equation*}
		|f(t, \xi)|\leq|t|^{-1/2}\|f\|_{\mathcal{Y}}
	\end{equation*}
	for all $t$ and $\xi$. Hence we can assume without any loss that $\|f\|_{\mathcal{Y}} = 1$ and we can focus
	on estimate of the following integral:
	\begin{equation*}
		I = |\xi|\int_{\mathbb{R}} e^{-|\xi||t-y|} |y|^{-1/2} dy.
	\end{equation*}
	We use again a substitution $u = \xi y$ to get:
	\begin{equation}\label{ch4:mainLem140}
		I = |\xi|^{1/2}\int_{\mathbb{R}}e^{-||\xi t| -|u||} u^{-1/2} du.
	\end{equation}
	We split this integral into three parts: $A_1 = \{u : |u| < |\xi||t|/2 \}$,
	$A_2 = \{u : |\xi||t|/2 \leq u \leq 2|\xi||t|\}$, 
	$A_3 = \{ u : 2|\xi||t| \leq u\}$ and we introduce $I_1$, $I_2$ and $I_3$ as the 
	corresponding parts of integral $I$.
	
	\noindent
	For $I_1$ we have:
	\begin{equation*}
	 I_1 \leq \frac{1}{2}|\xi|^{1/2}\int_{A_1} e^{-|\xi||t|/2} u^{-1/2} = |\xi|^{1/2} e^{-|\xi||t|/2} |\xi t|^{1/2}.
	\end{equation*}
	Now since
	\begin{equation}\label{ch4:mainLem150}
		|\xi t|e^{-|\xi t|} \leq C,
	\end{equation}
	where $C$ is independent of $\xi t$, we have
	\begin{equation*}
		I_1 \leq |\xi|^{1/2} \frac{|\xi t|^{1/2}}{(1+|\xi t|)} = |t|^{-1/2}\frac{|\xi t|}{(1+|\xi t|)} \leq |t|^{-1/2},
	\end{equation*}
	which is the desired estimate.

	To estimate integrals $I_2$ and $I_3$ one can repeat reasoning from the first part of this proof. It is sufficient
	to notice that previously for integrals $I_2$ and $I_3$ in the first step we were extracting term $(u_\infty+t\xi^2)^{-2\beta+1/2}$
	in front of these integrals. Here we do not have this term so all other estimates are the same.

	The proof of the lemma has been finished.
\end{Proof}

\begin{Lemma}\label{ch4:mainLem195}
	Let $f\in \mathcal{Y}$. Given the following terms:
	\begin{displaymath}
	\begin{array}{rclrcl}
\displaystyle		\tilde{A} & := & \frac{|\xi|}{\Delta}\int_{-\infty}^t e^{-\lambda_1(s-t)}f(s, \xi) ds,\qquad & \tilde{B} & := & \frac{\lambda_1}{\Delta}\int_{-\infty}^t e^{-\lambda_1(s-t)}f(s, \xi) ds,\\
		\displaystyle \tilde{C} & := & \frac{|\xi|}{\Delta}\int_{t}^\infty e^{-\lambda_2(s-t)}f(s, \xi) ds,\qquad & \tilde{D} & := & \frac{\lambda_2}{\Delta}\int_{t}^\infty e^{-\lambda_2(s-t)}f(s, \xi) ds.
	\end{array}
	\end{displaymath}
	Then $\tilde{A}, \tilde{B}, \tilde{C}, \tilde{D} \in \mathcal{X}_\beta$ provided $2\beta \leq 1$
	and the following estimate is valid:
	\begin{equation*}
		\|\tilde{A}\|_{\mathcal{X}_\beta} + \|\tilde{B}\|_{\mathcal{X}_\beta} + \|\tilde{C}\|_{\mathcal{X}_\beta} + \|\tilde{D}\|_{\mathcal{X}_\beta} 
			\leq c\|f\|_{\mathcal{Y}}.
	\end{equation*}
	\begin{equation*}
		\tilde{I} := \frac{|\xi|}{\Delta}\int_{-\infty}^t e^{-\lambda_1(s-t)}f(s, \xi) ds
	\end{equation*}
	belongs to the function space $\mathcal{X}_\beta$ provided $2\beta \leq 1$
	and the following estimate is valid:
	\begin{equation*}
		\|\tilde{I}\|_{\mathcal{X}_\beta}\leq \|f\|_{\mathcal{Y}}.
	\end{equation*}
	\textbf{Remark}: The same estimate is valid for
	\begin{equation*}
		\tilde{I} := \frac{\lambda_1}{\Delta}\int_{-\infty}^t e^{-\lambda_1(s-t)}f(s, \xi) ds.
	\end{equation*}	
\end{Lemma}

\begin{Proof}{}
	Let us start with integral $\tilde{A}$.
	Without loss of generality we may assume that $t > 0$.
	As in previous lemmas we may also assume that $\|f\|_{\mathcal{Y}} = 1$ and consider integral
	\begin{equation*}
		A := \frac{|\xi|}{\Delta}\int_{-\infty}^t e^{-\lambda_1(s-t)}|s|^{-1/2} ds.
	\end{equation*}
	Since $\lambda_1 = (u_\infty-\Delta)/2$ and $\Delta = \sqrt{u_\infty^2+4\xi^2}$ we see, that the behaviour of $\lambda_1$
	is different for small $\xi$ and large $\xi$. 
	
	Let us assume, that $|\xi| < u_\infty$. In this case we have $\Delta \sim u_\infty$ and $\lambda_1 \sim -|\xi|^2/u_\infty$, thus:
	\begin{equation*}
		A \leq C\frac{|\xi|}{u_\infty} \int_{-\infty}^t e^{|\xi|^2/u_\infty(s-t)}|s|^{-1/2} ds.
	\end{equation*}
	Using a substitution $-u = |\xi|^2(s-t)$ we get:
	\begin{equation*}
		A \leq C\frac{1}{u_\infty} \int_{0}^\infty e^{-u/u_\infty}|u-t\xi^2|^{-1/2} du = Cu_\infty^{-1/2}\int_0^\infty e^{-y} |y-t\xi^2|^{-1/2} dy.
	\end{equation*}
	We use inequality (\ref{ch4:mainIneq}) to obtain:
	\begin{equation}\label{ch4:mainLem310}
		A \leq Cu_{\infty}^{-1/2}(1+t\xi^2/u_\infty)^{-1/2} = C(u_\infty + t\xi^2)^{-1/2},
	\end{equation}
	which is the desired estimate, since for $2\beta \leq 1$ we have:
	\begin{equation*}
		(u_\infty + t\xi^2)^{\beta} A \leq (u_\infty + t\xi^2)^{\beta - 1/2} \leq c,
	\end{equation*}
	where constant $c$ does not depend on $t$, $\xi$ and $u_\infty$, thus $A \in \mathcal{X}_\beta$.

	Let us now assume that $|\xi| \geq u_\infty$. In this case we have $\lambda_1 \sim -|\xi|$ and $\Delta \sim |\xi|$, thus:
	\begin{equation}\label{ch4:mainLem185}
		A \leq C\int_{-\infty}^t e^{|\xi|(s-t)}|s|^{-1/2} ds.
	\end{equation}	
	Using a substitution $|\xi|s = -u + t|\xi|$ we end up with:
	\begin{multline*}
		A \leq  |\xi|^{-1/2}|\int_0^\infty e^{-u} |t|\xi|-u|^{-1/2}du 
		  \leq  |\xi|^{-1/2}(1+t|\xi|)^{-1/2} \\
			= (|\xi|+t|\xi|^2)^{-1/2} \leq  (u_\infty + t|\xi|^2)^{-1/2},
	\end{multline*}
	where we used again inequality (\ref{ch4:mainIneq}) with $\theta = t|\xi|$ and the fact, that $|\xi| > u_\infty$.
	This is the same inequality as (\ref{ch4:mainLem310}), thus we have proved results of Lemma \ref{ch4:mainLem195} for $\tilde{A}$.
	
	To prove estimate for $\tilde{B}$ we notice that $\lambda_1$ can be estimated by $|\xi|$, since for small $|\xi|$, i.e. $|\xi|\leq u_\infty$
	one has $\lambda_1 \sim |\xi|^2/u_\infty \leq |\xi|$, and for large $|\xi|$ one has $\lambda_1 \sim |\xi|$.

	To estimate $\tilde{C}$	we first notice, that without loss of generality we may assume $t < 0$, i.e. $t = -|t|$. 
	It is easy to see that we may show this inequality for $t < 0$, i.e. 
	As earlier -- the behaviour of $\lambda_2 = (u_\infty + \Delta)/2$ is different for small $|\xi|$ and large $|\xi|$, 
	in particular $\lambda_2 \sim u_\infty$ for small $|\xi|$ and $\lambda_2 \sim |\xi|$ for large $|\xi|$. Term $\Delta$ behave exactly the same.
	
	Let us first consider the case $|\xi|\leq u_\infty$. We have
	\begin{eqnarray*}
		C & = & \frac{|\xi|}{\Delta}\int_t^\infty e^{-\lambda_2(s-t)}|s|^{-1/2}ds 
		  \leq \int_t^\infty e^{-u_\infty(s-t)}|s|^{-1/2} ds\\
		  & = & u_\infty^{1/2} \int_0^\infty e^{-u_\infty u} |u-|t||^{-1/2} du 
		  = u_\infty^{-1/2} \int_0^\infty e^{-y}|y - u_\infty|t||^{-1/2}dy.
	\end{eqnarray*}
	We may now use (\ref{ch4:mainIneq}) to obtain:
	\begin{equation*}
		C \leq u_\infty^{-1/2} (1+u_\infty|t|)^{-1/2} = (u_\infty + |t|u_\infty^2)^{-1/2},
	\end{equation*}
	but since $|\xi| < u_\infty$ the last term can be estimated by $(u_\infty + |t||\xi|^2)^{-1/2}$, which, as we have seen earlier,
	is the desired estimate.
	
	Let us now assume that $|\xi| > u_\infty$. In this case we have:
	\begin{equation*}
		C = \frac{|\xi|}{\Delta}\int_t^\infty e^{-\lambda_2(s-t)}|s|^{-1/2}ds \leq \int_t^\infty e^{-|\xi|(s-t)}|s|^{-1/2} ds,
	\end{equation*}
	and the last term estimates exactly in the same was as in (\ref{ch4:mainLem185}), thus:
	\begin{equation*}
		C \leq C(u_\infty + t|\xi|^2)^{-1/2}.
	\end{equation*}

	Estimate of $\tilde{D}$ is analogous to the previous one, since for large $|\xi|$ both integrals
	behave exactly the same, and for small $|\xi|$ one has $\lambda_2 \sim u_\infty$, hence
	we must estimate integral:
	\begin{equation*}
		D = \frac{\lambda_2}{\Delta} \int_t^{\infty} e^{-u_\infty(s-t)}|s|^{-1/2}ds 
			\leq \int_t^{\infty} e^{-u_\infty(s-t)}|s|^{-1/2}ds,
	\end{equation*}
	but exactly the same integral has been estimated during the proof of estimate for $C$.
\end{Proof}

The previous Theorem needs to be modified in order to be used for investigation of the asymptotic profile:
\begin{Lemma}\label{new:mainLem195}
	Let $f$ satisfy the following pointwise inequality:
	\begin{equation}
		|f(t, \xi)| \leq |t|^{-1/2}(u_\infty + |t\xi|^2)^{-2\beta + \frac{1}{2}}.
	\end{equation}
	Then for $1/2\geq \beta > 1/4$ and for the given terms:
	\begin{displaymath}
	\begin{array}{rclrcl}
\displaystyle		\tilde{A} & := & \frac{|\xi|}{\Delta}\int_{-\infty}^t e^{-\lambda_1(s-t)}f(s, \xi) ds,\qquad & \tilde{B} & := & \frac{\lambda_1}{\Delta}\int_{-\infty}^t e^{-\lambda_1(s-t)}f(s, \xi) ds,\\
		\displaystyle \tilde{C} & := & \frac{|\xi|}{\Delta}\int_{t}^\infty e^{-\lambda_2(s-t)}f(s, \xi) ds,\qquad & \tilde{D} & := & \frac{\lambda_2}{\Delta}\int_{t}^\infty e^{-\lambda_2(s-t)}f(s, \xi) ds.
	\end{array}
	\end{displaymath}
	Then $\tilde{A}, \tilde{B}, \tilde{C}, \tilde{D} \in \mathcal{X}_{2\beta}$
	and the following estimate is valid:
	\begin{equation*}
		\|\tilde{A}\|_{\mathcal{X}_{2\beta}} + \|\tilde{B}\|_{\mathcal{X}_{2\beta}} + \|\tilde{C}\|_{\mathcal{X}_{2\beta}} + \|\tilde{D}\|_{\mathcal{X}_{2\beta}} 
			\leq C.
	\end{equation*}
\end{Lemma}

\begin{Proof}{}
	Let us start with integral $\tilde{A}$.
	Without loss of generality we may assume that $t > 0$ and consider the following integral:
	\begin{equation*}
		A := \frac{|\xi|}{\Delta}\int_{-\infty}^t e^{-\lambda_1(s-t)}|s|^{-1/2} (u_\infty+|s\xi^2|)^{-2\beta+1/2} ds.
	\end{equation*}
	Since $\lambda_1 = (u_\infty-\Delta)/2$ and $\Delta = \sqrt{u_\infty^2+4\xi^2}$ we see, that the behaviour of $\lambda_1$
	is different for small $\xi$ and large $\xi$. 
	
	Let us assume, that $|\xi| < u_\infty$. In this case we have $\Delta \sim u_\infty$ and $\lambda_1 \sim -|\xi|^2/u_\infty$, thus:
	\begin{equation*}
		A \leq C\frac{|\xi|}{u_\infty} \int_{-\infty}^t e^{|\xi|^2/u_\infty(s-t)}|s|^{-1/2}(u_\infty+|s\xi^2|)^{-2\beta+1/2} ds.
	\end{equation*}
	Using a substitution $-u = |\xi|^2(s-t)$ we get:
	\begin{multline*}
		A \leq C\frac{1}{u_\infty} \int_{0}^\infty e^{-u/u_\infty}|u-t\xi^2|^{-1/2} (u_\infty+|u-t\xi^2|)^{-2\beta+1/2}du = \\ 
			= Cu_\infty^{-2\beta+1/2-1/2}\int_0^\infty e^{-y} |y-t\xi^2/u_\infty|^{-1/2} (1+|y-t\xi^2/u_\infty|)^{-2\beta+1/2} dy.
	\end{multline*}
	We use inequality (\ref{new:mainIneq}) to obtain:
	\begin{equation}\label{new:mainLem310}
		A \leq Cu_{\infty}^{-2\beta}(1+t\xi^2/u_\infty)^{-2\beta} = C(u_\infty + t\xi^2)^{-2\beta},
	\end{equation}
	which is the desired estimate.

	Let us now assume that $|\xi| \geq u_\infty$. In this case we have $\lambda_1 \sim -|\xi|$ and $\Delta \sim |\xi|$, thus:
	\begin{equation}\label{new:mainLem185}
		A \leq C\int_{-\infty}^t e^{|\xi|(s-t)}|s|^{-1/2}(u_\infty+s\xi^2)^{-2\beta+1/2} ds.
	\end{equation}	
	Using a substitution $|\xi|s = -u + t|\xi|$ we end up with:
	\begin{eqnarray*}
		A & \leq & \int_0^\infty |\xi|^{-2\beta} e^{-u} |t|\xi|-u|^{-1/2}\left(\frac{u_\infty}{|\xi|} + |t|\xi|-u|\right)^{-2\beta+1/2}du \nonumber\\
		  & = & \int_{0}^{t\xi/2}+\int_{t\xi/2}^{2t\xi}+\int_{2t\xi}^\infty =: A_1 + A_2 + A_3.
	\end{eqnarray*}
	Now we estimate integral $A_1$, $A_2$ and $A_3$ separately.

	Integral $A_1$ can be estimated as follows:
	\begin{multline}
		A_1 = |\xi|^{-2\beta} \int_{0}^{t\xi/2} e^{-u}|t\xi|^{-1/2}\left(\frac{u_\infty}{|\xi|} + |t\xi|\right)^{-2\beta+1/2} =\\
			= |\xi|^{-2\beta}(1-e^{-t\xi/2})|t\xi|^{-1/2}\left(\frac{u_\infty}{|\xi|} + |t\xi|\right)^{-2\beta+1/2}.
	\end{multline}
	We now consider two cases: the first one is for $|t\xi| < u_\infty/|\xi|$. Remembering that $u_\infty/|\xi| \leq 1$ and that
	$(1-e^{-t\xi/2})|t\xi|^{-1/2} \sim |t\xi|^{1/2}$ we estimate $A_1$:
	\begin{multline}
		A_1 \leq |\xi|^{-2\beta}|t\xi|^{1/2}\left(\frac{u_\infty}{|\xi|} + |t\xi|\right)^{-2\beta+1/2} 
			\leq |\xi|^{-2\beta}\left(\frac{u_\infty}{|\xi|} + |t\xi|\right)^{-2\beta+1/2+1/2} \\
			\leq \left(u_\infty + t\xi^2\right)^{-2\beta},
	\end{multline}
	where we used the same assumption $|t\xi| < u_\infty/|\xi|$.

	In case $t|\xi| > u_\infty /|\xi|$ we may write $|t\xi|^{-1/2} \leq (|t\xi| + u_\infty/|\xi|)^{-1/2}$ which gives
	us the following easy estimate:
	\begin{equation}
		A_1 \leq |\xi|^{-2\beta}\left(\frac{u_\infty}{|\xi|} + |t\xi|\right)^{-2\beta+1/2-1/2} = (u_\infty+t\xi^2)^{-2\beta}.
	\end{equation}

	Let us now estimate integral $A_2$. We first estimate it in the following way:
	\begin{equation}
	 	A_2 \leq \int_{0}^{|t\xi|}|\xi|^{-2\beta}e^{-|t\xi|}|u|^{-1/2}\left(\frac{u_\infty}{|\xi|} + |u|\right)^{-2\beta+1/2} du
	\end{equation}
	
	Now for the first case $|t\xi| < u_\infty/|\xi|$ one can write:
	\begin{equation}
		A_2 \leq |\xi|^{-2\beta}e^{-t|\xi|} |t\xi|^{1/2} \left(\frac{u_\infty}{|\xi|}\right)^{-2\beta+1/2} 
			\leq |\xi|^{-2\beta}\left(\frac{u_\infty}{|\xi|}\right)^{-2\beta+1} \leq u_\infty^{-2\beta},
	\end{equation}
	which under our assumption is the desired estimate since $u_\infty^{-2\beta} \leq (u_\infty+t\xi^2)^{-2\beta}$.

	Let us now consider the case $|t\xi| > u_\infty/|\xi|$. For this we split integral $A_2$ as $A_2 = \int_{0}^{u_\infty/|\xi|} + \int_{u_\infty/|\xi|}^{|t\xi|} =: A_{21} + A_{22}$.

	Integral $A_{21}$ we estimate as follows:
	\begin{equation}
	 A_{21} \leq \int_{0}^{u_\infty/|\xi|} |\xi|^{-2\beta}e^{-t|\xi|} |u|^{-1/2}({u_\infty/|\xi|})^{-2\beta+1/2} du =
		u_\infty^{-2\beta} e^{-t\xi}|{u_\infty/|\xi|}|^{1}.
	\end{equation}
	We estimate $e^{-t|\xi|} \leq (1+t|\xi|)^{-1}$ and get
	\begin{equation}
	 A_{21} \leq u_\infty^{-2\beta+1}(|\xi|+t\xi^2)^{-1} \leq u_\infty^{-2\beta+1}(u_\infty+t\xi^2)^{-1},
	\end{equation}
	where we again used $|\xi|>u_\infty$. Now if $\beta \leq 1/2$ (i.e. $-1 \leq -2\beta$) one estimates the last
	term by $u_\infty^{-2\beta+1}(u_\infty+t\xi^2)^{-2\beta}$.

	For integral $A_3$ one changes variables $-y = t|\xi| - u$ to get:
	\begin{multline}
	  A_3 = \int_{t\xi}^{\infty}|\xi|^{-2\beta}e^{-t\xi} e^{-y}|y|^{-1/2}\left(\frac{u_\infty}{|\xi|} + |y|\right)^{-2\beta+1/2}dy \\
		\leq e^{-t|\xi|}|\xi|^{-2\beta}\left(\frac{u_\infty}{|\xi|} + |t\xi|\right)^{-2\beta+1/2}\int_{0}^{\infty}e^{-y}|y|^{-1/2}dy.
	\end{multline}
	Under assumption that $u_\infty \leq |\xi|$ one easily gets:
	\begin{equation}
	 A_3 \leq C(1+|t\xi|)^{1/2}e^{-t|\xi|}\left(u_\infty + t\xi^2\right)^{-2\beta} \leq C\left(u_\infty + t\xi^2\right)^{-2\beta}.
	\end{equation}
	This finishes the estimate for the term $A$.

	As in the proof of Lemma \ref{ch4:mainLem195}: to prove estimate for $\tilde{B}$ we notice that $\lambda_1$ can be estimated by $|\xi|$, since for small $|\xi|$, i.e. $|\xi|\leq u_\infty$
	one has $\lambda_1 \sim |\xi|^2/u_\infty \leq |\xi|$, and for large $|\xi|$ one has $\lambda_1 \sim |\xi|$. Thus $B < A$.
	
	To estimate $C$ we first notice that without loss of generality we may assume $t < 0$. Using the identity $\lambda_1 + \lambda_2 = u_\infty$ we may estimate integral $C$ as follows:
	\begin{multline}
	 	C(t, \xi) \leq \frac{|\xi|}{\Delta}\int_{t}^\infty e^{-\lambda_2(s-t)}|f(s, \xi)|ds = \\
			\frac{|\xi|}{\Delta}\int_{t}^\infty e^{\lambda_1(s-t)}e^{-u_\infty(s-t)}|f(s, \xi)|ds\leq
			\frac{|\xi|}{\Delta}\int_{t}^\infty e^{\lambda_1(s-t)}|f(s, \xi)|ds.
	\end{multline}
	Next using a substitution $s = -u$ and denoting $\tau = -t$ we get:
	\begin{equation}
	 C(-\tau, \xi) \leq \frac{|\xi|}{\Delta}\int_{-\infty}^\tau e^{-\lambda_1(u-\tau)}|f(-u, \xi)|du,
	\end{equation}
	but this integral has been already estimated above (that is in case of integral $A$), one just needs to remember
	that the estimate for $|f|$ does not depend on the sign of $u$. 

	To complete the proof of this theorem we need to estimate integral $D$.  First we notice that for large $|\xi|$, that
	is for $|\xi| > u_\infty$ integral $D$ and $C$ behave exactly the same, thus we estimate $D$ only under condition $|\xi| < u_\infty$. 
	Then $\lambda_2 \sim u_\infty$ and $\Delta \sim u_\infty$.

	As in case of integral $C$ we also assume without loss of generality that $t < 0$. For convenience let us denote $-\gamma = -2\beta + 1/2$. 
	We then have:
	\begin{multline}
		D \leq \int_{t}^\infty e^{-u_\infty(s-t)} |s|^{-1/2}(u_\infty+s\xi^2)^{-\gamma} ds = \\
			= \int_{0}^\infty e^{-u}|u_\infty|^{-1/2}|u+u_\infty t|^{-1/2}\left(u_\infty+|u+u_\infty t|\frac{\xi^2}{u_\infty}\right)^{-\gamma} du =\\
			= \int_{0}^{tu_\infty/2} + \int_{tu_\infty/2}^{2tu_\infty} + \int_{2tu_\infty}^{\infty} =: D_1 + D_2 + D_3,
	\end{multline}
	where we use a substitution $u_\infty(s-t) = u$.

	We estimate integral $D_1$ as follows:
	\begin{multline}
	 D_1 \leq \int_{0}^{tu_\infty/2} e^{-u}|u_\infty|^{-1/2}|tu_\infty|^{-1/2}\left(u_\infty+tu_\infty\frac{\xi^2}{u_\infty}\right)^{-\gamma} = \\
		= (1-e^{-tu_\infty/2})|u_\infty|^{-1/2}|tu_\infty|^{-1/2}(u_\infty+t\xi^2)^{-\gamma} \\
		\leq \frac{|tu_\infty|^{1/2}}{(1+|tu_\infty|)} |u_\infty|^{-1/2}(u_\infty+|t\xi^2|)^{-\gamma}
		\leq (u_\infty+|tu_\infty^2|)^{-1/2}(u_\infty+|t\xi^2|)^{-\gamma}.
	\end{multline}
	But now since $|\xi|<u_\infty$ one has $(u_\infty+|tu_\infty^2|)^{-1/2} \leq (u_\infty+|t\xi^2|)^{-1/2}$ and thus:
	\begin{equation}
	 D_1 \leq (u_\infty+|t\xi^2|)^{-\gamma - 1/2} = (u_\infty+|t\xi^2|)^{-2\beta},
	\end{equation}
	which is the desired estimate.

	For integral $D_2$ we proceed as follows:
	\begin{equation}
	 D_2 = \int_{tu_\infty/2}^{2tu_\infty} \leq e^{-tu_\infty/2}|u_\infty|^{-1/2}\int_{0}^{tu_\infty} |u|^{-1/2}\left(u_\infty+u\frac{\xi^2}{u_\infty}\right)^{-\gamma}du.
	\end{equation}
	Now we distinguish two cases. The first one is for $|t\xi^2| < u_\infty$, that is for $tu_\infty\frac{\xi^2}{u_\infty} < u_\infty$. Then we may estimate $D_2$ as:
	\begin{equation}
	 D_2 \leq e^{-tu_\infty/2}|u_\infty|^{-1/2}|tu_\infty|^{1/2}u_\infty^{-\gamma} \leq
		u_\infty^{-\gamma-1/2} \leq (u_\infty + |t\xi^2|)^{-\gamma-1/2},
	\end{equation}
	where we used the fact that $e^{-tu_\infty/2}|tu_\infty|^{1/2}$ is uniformly bounded.

	For $|t\xi^2| \geq u_\infty$ one has:
	\begin{multline}
	 D_2 \leq e^{-tu_\infty/2}|u_\infty|^{-1/2}\int_{0}^{tu_\infty}|u|^{-1/2}|u|^{-\gamma}|\xi^2|^{-\gamma}u_\infty^\gamma du = \\
		= e^{-tu_\infty/2}|u_\infty|^{-1/2}|tu_\infty|^{-1/2-\gamma+1}|\xi^2|^{-\gamma}u_\infty^{\gamma} \\
			\leq |t\xi^2|^{-\gamma-1/2}u_\infty^{1/2}.
	\end{multline}
	where we again used assumption $|\xi| < u_\infty$ and assumption $\gamma < 1/2$ which corresponds to $\beta < 1/2$. The last inequality
	under our assumption of $|t\xi^2| \geq u_\infty$ gives us the desired estimate:
	\begin{equation}
	 D_2 \leq (u_\infty+|t\xi^2|)^{-2\beta}.
	\end{equation}

	To finish the proof of the Lemma we need to estimate integral $D_3$. First we get:
	\begin{multline}
	 D_3 = \int_{2tu_\infty}^{\infty} \leq \int_{tu_\infty}^{\infty}e^{-u}|u_\infty|^{-1/2}|u|^{-1/2}\left(u_\infty+|u|\frac{\xi^2}{u_\infty}\right)^{-\gamma} \\
			\leq (u_\infty + |t\xi|^2)^{-\gamma}\int_{tu_\infty}^{\infty} e^{-u}|u_\infty|^{-1/2}|u|^{-1/2}.
	\end{multline}
	Now in case of $t\xi^2 < u_\infty$ one easily gets:
	\begin{equation}
	 D_3 \leq (u_\infty + |t\xi^2|)^{-\gamma} u_\infty^{-1/2}\int_{0}^\infty e^{-u}|u|^{-1/2} \leq (u_\infty+t\xi^2)^{-\gamma-1/2},
	\end{equation}
	which is the desired estimate. In case $t\xi^2 \geq u_\infty$ one first notice that this condition implies
	$tu_\infty^2 > t\xi^2 > u_\infty$, that is $tu_\infty > 1$. This allows us to estimate $D_3$ as follows:
	\begin{multline}
	 D_3 \leq (u_\infty + |t\xi^2|)^{-\gamma}(2tu_\infty)^{-1/2} u_\infty^{-1/2} \int_0^\infty e^{-u} du \\
		\leq (u_\infty + |t\xi^2|)^{-\gamma}(1+tu_\infty)^{-1/2}u_\infty^{-1/2}
		\leq (u_\infty + |t\xi^2|)^{-\gamma}(u_\infty+|tu_\infty^2|)^{-1/2} \\
		\leq (u_\infty + |t\xi^2|)^{-\gamma-1/2},
	\end{multline}
	where again we used assumption $|\xi| < u_\infty$. 

	This completes the proof of Lemma \ref{new:mainLem195}.

\end{Proof}

% section Main Lemmas (end)

\textsl{Acknowledgement.} 
Authors have been supported by Polish grant No. N N201 547 438.

\end{document}